\documentclass[preprint2]{aastex}

\shortauthors{Burgasser et al.}
\shorttitle{Variability in 2MASS 1237+6526}

\begin{document}

\title{A Search for Variability in the Active T dwarf 2MASS 1237+6526}

\author{
Adam J.\ Burgasser\altaffilmark{1,2},
James Liebert\altaffilmark{3},
J.\ Davy Kirkpatrick\altaffilmark{4},
and John E.\ Gizis\altaffilmark{5}}
 
\altaffiltext{1}{Division of Physics \& Astronomy, University of California
at Los Angeles, Los Angeles,
CA, 90095-1562; adam@astro.ucla.edu}
\altaffiltext{2}{Hubble Postdoctoral Fellow}
\altaffiltext{3}{Steward Observatory, University of Arizona,
Tucson, AZ 85721; liebert@as.arizona.edu}
\altaffiltext{4}{Infrared Processing and Analysis Center, M/S 100-22, 
California Institute of Technology, Pasadena, CA 91125; davy@ipac.caltech.edu}
\altaffiltext{5}{Sharp Lab, Department of Physics \& Astronomy,
University of Delaware, Newark, DE 19716; gizis@udel.edu}

\begin{abstract}
We present spectroscopic and imaging observations of the 
active T dwarf 2MASS 1237+6526, intended to investigate the emission
mechanism of this cool brown dwarf.
The H$\alpha$ emission line first detected
in 1999 July appears to be persistent over 1.6 years, 
with no significant variation
from $\log_{10}$(L$_{H{\alpha}}$/L$_{bol}$) = $-$4.3,
ruling out flaring as a possible source.
The relatively high
level of emission in this object
appears to be unique amongst observed late-L and 
T dwarfs.
One of our spectra shows an apparent velocity shift 
in the H$\alpha$ line, which could be indicative of an accretion hot spot
in orbit around the brown dwarf;
further confirmation of this shift is required.
J-band monitoring observations fail to detect any significant variability
(e.g., eclipsing events)
at the $\pm$0.025 mag level over periods of up to 2.5 hours, and there 
appears to be no statistical evidence of
variability for periods of up to 14 hours.
These limits constrain the mass of a hypothetical interacting secondary
to M$_2$ $\lesssim$ 20 M$_{Jup}$ for inclinations $i$ $\gtrsim$ 60$\degr$.
While our observations do not explicitly rule out the binary hypothesis
for this object, it does suggest that other mechanisms, such as youthful
accretion, may be responsible.
\end{abstract}


\section{Introduction}

A previously undetected population of 
T dwarfs, brown dwarfs exhibiting CH$_4$ absorption 
bands at 1.6 and 2.2
$\micron$ \citep{kir99,me01a}, have recently been identified by various 
deep optical and near-infrared surveys 
\citep{str99,me99,me00a,me00c,me01a,cub99,tsv00,leg00,geb01b}.  One of these
objects,  
2MASSI J1237392+652615
\citep[hereafter, 2MASS 1237+6526]{me99}, 
identified in the Two Micron All Sky Survey 
\citep[hereafter 2MASS]{skr97}, was found to have H$\alpha$ in
emission, with a relative luminosity 
$\log_{10}$(L$_{H{\alpha}}$/L$_{bol}$) = $-$4.3 \citep{me00b}.
While many low-mass stars and young brown dwarfs are seen to
exhibit this feature, likely generated by
chromospheric magnetic activity, 
the frequency and relative luminosity of
H$\alpha$ emission drops rapidly beyond spectral types M7 V \citep{kir00,giz00}.
Indeed, at the time of its detection, the emission of 2MASS 1237+6526 was 
unique amongst objects later than type L5 V. 
\citet{kir01} have recently
reported weak H$\alpha$ emission in the bright T2 V 
SDSSp J125453.90-012247.4 \citep[hereafter SDSS 1254-0122]{leg00}.

In order to explain the activity of 2MASS 1237+6526,
\citet{me00b} have hypothesized that this object may actually be
an interacting brown dwarf binary system.  In
this scenario, the binary mass ratio and orbital separation are such that the
lower-mass secondary fills its Roche lobe and steadily loses mass to the primary.
For a 70 M$_{Jup}$\footnote{We adopt the definitions 
M$_{Jup}$ = 1.90$\times$10$^{30}$ g = 9.55$\times$10$^{-4}$ M$_{\sun}$, 
R$_{Jup}$ = 7.15$\times$10$^{9}$ cm = 0.10 R$_{\sun}$ \citep{cox00,tho00}.}
primary and mass ratio $q$ $\equiv$ $M_2/M_1$ $>$ 0.07 (i.e., $M_2$ $>$ 
5 M$_{Jup}$), sustained
mass loss can occur for physical separations $a$ $\lesssim$ 6
R$_{Jup}$ and orbital periods $p$ $\lesssim$ 5.5 hours.  The close orbit
required for this scenario currently rules out direct imaging of the binary
pair ($a$ $<$ 0$\farcs$0002 
assuming a distance of 14 pc; Burgasser et al.\ 1999);
however, partial eclipsing could be observed in a system with an 
inclination $i$ $\gtrsim$ 60$\degr$.  

We have obtained followup spectroscopic
and photometric observations of 2MASS 1237+6526 in order to test the
binary hypothesis, as well as examine
the possibility that this object was originally 
observed during a prolonged flare.  
In $\S$2, we present
red optical (6300--10100 {\AA}) spectral data obtained over 1.6 yr
using the Keck 10m 
Low Resolution Imaging Spectrograph \citep[hereafter LRIS]{oke95},
and discuss the behavior of the H$\alpha$
emission line over this period.
J-band monitoring observations obtained over two nights using the
Palomar 60'' Near-Infrared Camera
\citep[hereafter IRCam]{mur95} are presented in $\S$3.
We discuss how the imaging observations constrain the 
binary scenario in $\S$4.  Finally, in $\S$5 we discuss how these
follow-up observations constrain various emission mechanisms for
this unique object.

\section{Spectroscopic Observations}

2MASS 1237+6526 has been observed on seven separate occasions 
in the red optical regime starting 
in 1999 July.  Early data, obtained on 1999 July 16-18 (UT) using LRIS,
are described in  \citet{me00b}.  
Here, we present 
our latest LRIS observations obtained on 2000 March 5 and 2001 February 20 (UT).
Data acquisition and reduction procedures for 
the first run are described below, and are similar to those used for the
2001 February data as described in \citet{kir01}. 

\subsection{Data Acquisition and Reduction}

Conditions during our 2000 March 5 (UT) observations of
2MASS 1237+6526 were clear with 0$\farcs$8--0$\farcs$9 seeing.  
Because of its optical faintness (R $\gtrsim$ 26; 
Matthews et al.\ 1996, Burgasser et al.\ 1999), the target was invisible
on the guider camera, so acquisition into a 1$\arcsec$ slit was done
by blind offset from a nearby optical source using positions
determined from 2MASS coordinates.  Spectral data were
obtained using the 400 lines mm$^{-1}$ grating blazed at 8500 {\AA}, 
covering the spectral range 6300 to 10100 {\AA} at 9 {\AA} resolution.
The OG570 order blocking filter was used to reduce second-order light.
Two exposures of 1800s each were obtained; however, the data
from the first exposure were fainter than expected, so the target was 
repositioned prior to the second exposure.  
Observations of the DZA4 white dwarf
WD 1225-079 \citep{gre86,kil86} and the B1 V
flux standard Hiltner 600 \citep{hil56,ham94}
were also obtained for calibration using the same instrumental configuration.

Science data were reduced using 
standard IRAF\footnote{Image Reduction and Analysis Facility (IRAF) 
is distributed 
by the National Optical Astronomy Observatories,
which are operated by the Association of Universities for Research
in Astronomy, Inc., under cooperative agreement with the National
Science Foundation.} routines. 
A series of 
one-second dark exposures were median-combined
and subtracted from the science data to remove detector bias.   
Median-combined quartz-lamp 
flat-field exposures (reflected off of the interior dome)
were used to normalize the
response of the detector. Spectra were then extracted 
using the APEXTRACT routine.  Curvature of the dispersion line
was determined from bright standard star exposures and applied
to the extraction of the 2MASS 1237+6526 spectrum.
Wavelength calibration was done using Ne-Ar arc lamp exposures taken 
immediately after the target observations.  We then computed a telluric 
correction from the white dwarf spectrum by interpolating over
telluric O$_2$ (6850--6900 {\AA} A-band, 7600--7700 {\AA} B-band) and
H$_2$O (7150--7300, 8100--8350, and 9300--9650 {\AA}) bands,
and we applied this correction to our flux standard and 2MASS 1237+6526
spectra\footnote{Despite the presence of Ca II HK absorption in
this object \citep{kil86,lie87}, the
triplet lines at 8498, 8542, and 8662 {\AA} were not present
and hence did not affect the telluric correction.}.  
Finally, flux calibration was done by correcting our
Hiltner 600 observations to the spectrophotometric data given by
\citet{ham94}, being careful to interpolate over 
telluric features and the Balmer H$\alpha$ line.
This correction was then applied to the spectrum of 2MASS 1237+6526.

\placefigure{fig-1}

Reduced spectral data from our 2000 March and 2001 February 
observations are shown in Figure 1.  Full spectra with flux plotted
on a logarithmic scale are shown in the right panel, 
while a close up of
the H$\alpha$ line on a linear scale is shown on the left.  
A combined spectrum from our
1999 July 16 (UT) observations is also plotted for comparison \citep{me00b}.
The 2000 March spectrum is roughly 2.6 times
fainter than our 1999 July observations, likely the result of 
poor targeting onto the slit due to the blind offset and proper motion
of 2MASS 1237+6526  (first observed by 2MASS on 1999 March 13 UT).
The 2001 February spectrum, on the other hand, is brighter by a factor
of 1.3; atmospheric transparency on this night was excellent, and the target
was visible on the facility guide camera, enabling proper slit
placement.  Features present in the spectra are noted, including the 
strong 3(${\nu}_1$,${\nu}_3$) and 2($\nu$$_1$,$\nu$$_3$)+2$\nu$$_2$
H$_2$O bands starting at 
9250 and 9540 {\AA} \citep{aum67},
a possibly weak
4(${\nu}_1$,${\nu}_3$) CH$_4$ band centered at 8950 {\AA} \citep{dic77},
and lines of Cs I at 8521 and 8943 {\AA} 
($6s ^2S_{1/2} {\rightarrow} 6p ^2P_{1/2,3/2}$).
The dominant feature in these spectra is the strongly pressure-broadened
K I resonance doublet 
($4s ^2S_{1/2} {\rightarrow} 4p ^2P_{1/2,3/2}$), centered
at 7665 and 7699 {\AA}, but extending over most of the spectral regime
shown.  \citet{lie00} have shown that the red wing of the 
K I doublet 
is the primary contributor to
the steep slope seen from 8000--10000 {\AA} in the spectrum
of SDSSp J162414.37+002915.6 \citep{str99}, and this feature is likely 
responsible for shaping the red optical spectrum of 2MASS 1237+6526 as well.

\subsection{H$\alpha$ Emission in 2MASS 1237+6526}

H$\alpha$ emission at 6563 {\AA} is clearly seen in 
all of the spectra of Figure 1, 
and its presence in our
recent data suggests that emission is persistent in  
2MASS 1237+6526 over a period of at least 1.6 years.  
We have investigated the temporal behavior 
of H$\alpha$ emission by measuring the 
line flux in all of our spectra to date.  Flux data were integrated
within 20 {\AA} of the emission peak after subtracting off the neighboring
pseudo-continuum, the scatter of which was used to estimate 
uncertainties.  Corrections have
also been computed to compensate for slit losses, by scaling
the individual spectra to match data from 1999 July 16 (UT) in the
8000--9500 {\AA} region.  The initial and corrected line fluxes
are listed in Table 1. 

\placetable{tbl-1}
\placefigure{fig-2}

Figure 2 plots these line fluxes versus UT time.   All of the 
corrected values are in excellent agreement with the mean
line flux from our 1999 July observations, 
(7.0$\pm$0.5)$\times$10$^{-17}$ erg cm$^{-2}$ s$^{-1}$,
with the exception of data taken on 1999 July 17 (UT).
As reported by \citet{me00b}, these data were of extremely low
signal-to-noise (again, likely due to poor slit placement), 
and it is possible that the formal errors are too low.
Excepting this data point, we find that the
H$\alpha$ emission in 2MASS 1237+6526 is very stable,
firmly ruling out
any kind of flaring event.
The weighted mean value over all observations
is $f_{H{\alpha}}$ = (7.44$\pm$0.08)$\times$10$^{-17}$ erg cm$^{-2}$ s$^{-1}$,
consistent with previous results.

\subsection{Comparison to Other Cool Stars and Brown Dwarfs}

To gauge how the strength of H$\alpha$ emission in 
2MASS 1237+6526 compares to
other active, cool dwarfs, we computed the relative H$\alpha$ luminosity
for each spectrum, 
${\log}_{10}$(L$_{H\alpha}$/L$_{bol}$) = ${\log}_{10}$($f_{H\alpha}$/$f_{bol}$), 
where
\begin{equation}
{\log}_{10}f_{bol} = -0.4m_{bol} - 4.61
\end{equation}
in units of erg cm$^{-2}$ s$^{-1}$ \citep{dri00}, 
and $m_{bol}$ = $m_J$ + BC$_J$ is the apparent bolometric
magnitude.  Using a revised J-band magnitude for 2MASS 1237+6526 
from the 2MASS
Second Incremental Release Catalog
and BC$_J$ = 2.09$\pm$0.10 \citep[determined from measurements of
Gliese 229B\footnote{Gliese 229B and 2MASS 1237+6526 have the same
spectral type, T6.5 V \citep{me01a}; hence, differences in the bolometric
correction are likely to be negligible.}]{leg99}, 
we derive $m_{bol}$ = 18.12$\pm$0.14 and a mean 
${\log}_{10}$(L$_{H\alpha}$/L$_{bol}$) = $-$4.3,
consistent with the results
of \citet{me00b}.  

\placefigure{fig-3}

We compare this value to the emission ratios
of late-M, L, and T dwarfs in Figure 3.
Using spectral
data from \citet{kir99,kir00,kir01} and \citet{me00b,me01b}, we computed
H$\alpha$ line fluxes directly from the spectra as described above, 
with bolometric fluxes for late-M and L dwarfs
computed from 2MASS J-band photometry and BC$_J$ values from \citet{rei01}.
For T dwarfs, we used 2MASS J-band photometry and 
BC$_J$ = 2.09, yielding
an upper limit to the true bolometric correction
for the early- and mid-T dwarfs \citep{me01a},  
and hence a lower limit to the bolometric flux; i.e., overestimating the
relative H$\alpha$ luminosity.
Detections are plotted as
open symbols, while upper limits (determined from pseudocontinuum noise levels)
are indicated by downward arrows.  For comparison, we have included M and L dwarf
samples from \citet[crosses]{haw96} and \citet[diamonds]{giz00}.  The trend
of decreasing activity beyond M5--M7 V reported by \citet{giz00} is readily
apparent in this plot, and appears to continue well into the
T dwarf regime.  Indeed, only two
objects have reliable detections beyond spectral type L5 V, the T2 V
SDSS 1254-0122, with ${\log}_{10}$(L$_{H\alpha}$/L$_{bol}$) = $-$5.7, and 
2MASS 1237+6526. The former object is over 25 times less luminous in H$\alpha$
then 2MASS 1237+6526, and upper limits for the other T dwarfs do not
exceed ${\log}_{10}$(L$_{H\alpha}$/L$_{bol}$) = $-$5.2.

\citet{me00b} compares 2MASS 1237+6526 to another unusually
active cool dwarf, the M9.5 Ve PC 0025+0047 \citep{sch91}, which is indicated
in Figure 3 by the solid square.  This object, with
${\log}_{10}$(L$_{H\alpha}$/L$_{bol}$) = $-$3.4, is also more than
ten times brighter in H$\alpha$ than other objects in its class,
although it is not extremely active in comparison to
earlier emission M dwarfs \citep{haw96}.  PC 0025+0047
also shows long-term stability in its emission
\citep{sch91,mou94,mrt99}.  These common
characteristics provide further
evidence that emission processes in
these two objects may be related.

\subsection{A Possible Velocity Shift in the 2000 March Spectrum?}

Examination of the spectra of Figure 1 also shows an
apparent blueshift in the H$\alpha$ line in our 2000 March data 
as compared to the other epochs.  We measured a line center deviation of 
3.2$\pm$0.5 {\AA} in this spectrum with respect to the 2001 February data, 
where the uncertainty is derived from the
scatter of telluric OH emission line wavelengths between these two
data sets.  This deviation
corresponds to a velocity of 150$\pm$25 km s$^{-1}$,
which cannot be due to differences in 
the Earth's rotational and orbital motion between the two epochs, 
amounting to only 2.4 
km s$^{-1}$.

\placefigure{fig-4}

As with our initial detection of H$\alpha$,
we carefully scrutinized the reality of this subtle (1--2 pixel) line shift.
Figure 4 shows the H$\alpha$ region in the raw image data from 2000 March
and 2001 February.  Telluric OH emission lines in the 6563 {\AA}
region appear as vertical bands, 
while the bright spots are H$\alpha$ emission in 2MASS 1237+6526; note
the lack of continuum in this region.
There is a slight offset in the emission between these
two epochs with respect to the telluric lines, confirming the spectral
results in Figure 1.  
However, this shift could be due to poor centering of the
object on the slit. Both individual exposures taken on 2000 March
5 (UT) show a similar offset, despite repositioning of the object. 
The second exposure was brighter and probably better positioned,
but was still 2.6 times fainter
than our 1999 July results; hence, it too 
may have been misaligned.  
On the other hand, 
seeing on the nights of 2000 March 5 and 2001 February 20 (UT)
was roughly equivalent to the width of the slit employed, so
significant shifting of the
spectrum due to poor slit placement is less likely.  
We were unable to perform an independent check
using the weak 8521 and 8943 {\AA} Cs I lines  
due to the low signal-to-noise of both spectra. 
We therefore conclude that there is some
evidence for a velocity shift in the H$\alpha$ line which must be 
confirmed with higher resolution data.

Assuming its reality, the H$\alpha$ line shift (which appears to be
asymmetric toward the blue) could arise from a single hot spot on the surface
of a rapidly rotating brown dwarf.  The velocity of the line is 
comparable to but less than the break-up speed for a substellar object,
roughly 330 (520) km s$^{-1}$ for a 30 (75) M$_{Jup}$ brown dwarf. 
In this case,
however, the integration time of each individual
observation made (typically 20--30 minutes)
would be a significant fraction
of the rotation period (no less than about 50 minutes), and one would expect
the line to be broadened rather than shifted. 
A more reasonable source would be
a hot spot in orbit around the brown dwarf, possibly arising from an 
optically thin accretion
disk.  Discrete regions in an edge-on Keplerian disk would have
line-of-sight radial velocities as high as 150 km s$^{-1}$ at 
separations of roughly 5 (12) R$_{Jup}$ for a 30 (75) M$_{Jup}$ central
source.  The hot spot would have to be small, due to the lack of
blue continuum seen in at shorter wavelengths \citep{me00b}.
It is worth noting that the magnitude and direction of the 
velocity shift is similar to that seen in some solar flaring events 
during an impulsive evaporation phase \citep{ant85}; however, we do not see
the typically brighter redshifted condensation component \citep{fis85}.

\section{Imaging Observations}

In addition to our spectroscopic data, we searched for photometric
variability in 2MASS 1237+6526, for the purpose
of detecting an eclipse by a closely-separated companion.  

\subsection{Data Acquisition and Reduction}

J-band monitoring observations
were obtained over two nights, 2000 May 19--20
(UT), using the Palomar 60'' IRCam.
Conditions on both nights were generally hazy
but clear, with average seeing of 1$\farcs$2 to 1$\farcs$8; some
scattered cirrus was present during the observations of May 20.  
Data were acquired in sets of five 45s exposures dithered by 
10$\arcsec$ on May 19 and 15$\arcsec$ on May 20 in a square pattern.
A total of 155 and 165 images were obtained, 
spanning periods of 2.50 and 2.57 hours on May 19 and 20, respectively
(Table 2).   

\placetable{tbl-2}
\placefigure{fig-5}

Science data were initially 
pairwise differenced to eliminate bias and sky background,
and then divided by a normalized 
flat-field image, constructed by differencing median-combined
sets of lamp-on and lamp-off dome observations made on May 19.
Bad pixels were identified in the flat-field image and
corrected in the science images
by interpolation.  The four separate readouts of the IRCam
NICMOS-3 HgCdTe array occasionally introduced additive quadrant
biases that we removed by subtracting off the median
from each quadrant.  Finally, individual
images in each set were shifted and coadded to produce a final reduced
image (Figure 5).

\subsection{Differential Photometry and Analysis of the Time Series}

For each image, we computed
aperture photometry ($f_i$) for 2MASS 1237+6526 and
other sources in the field.  Flux within a
2$\farcs$5 (4.04 pixel) radius aperture was
integrated after subtracting the mean background measured
in a 6 to 12$\arcsec$ annulus around each object.
Uncertainty estimates of the photometry, ${\delta}f_i$,
included contributions by
shot noise in the source
and variations in the background (essentially flat field uncertainties). 
Instrumental magnitudes,
$m_i = -2.5{\log}_{10}f_i$, were then computed with corresponding
uncertainties, ${\sigma}_i = 1.086(\frac{{\delta}f_i}{f_i})$.  
The scatter in instrumental magnitudes
for 2MASS 1237+6526 was roughly 0.03 mag
on May 19 and 0.01 mag on May 20, with formal photometric
uncertainties of $\pm$0.025 mag.

To compute a relative magnitude light curve, we first examined 
sources in the field for variability, using the procedure outlined in 
\citet{bai01}.  We first
determined relative magnitudes and uncertainties for each source: 
\begin{equation}
\hat{m}_i^{(j)} = m_i^{(k)} - m_i^{ref}
\end{equation}
\begin{equation}
\hat{\sigma}_i^{{(j)}^2} = {{\sigma}_i^{(j)}}^2 + {{\sigma}_i^{ref}}^2,
\end{equation}
where the reference magnitudes and uncertainties were computed as
\begin{equation}
m_i^{ref} = -2.5{\log}_{10} \left( \sum_{k{\neq}j}^K f_i^{(k)} \right)
\end{equation}
\begin{equation}
{{\sigma}_i^{ref}}^2 = \sum_{k{\neq}j}^K {{\sigma}_i^{(k)}}^2,
\end{equation}
summing over the $K$ other sources in the field. 
After subtracting off the mean relative
magnitude, we computed the 
reduced ${\chi}^2$ for each source to 
determine if photometric variations were significant:
\begin{equation}
{\chi}_r^2 = \frac{1}{n-1} \sum_i^n (\frac{\hat{m}_i-{\langle}\hat{m}_i{\rangle}}{\hat{\sigma}_i})^2;
\end{equation}
here, $n$ is the total number of images.
We constrained calibrator sources to have
${\chi}_r^2$ $\lesssim$ 0.5, corresponding to a probability of
variability $P$ $\leq$ 1\% for $n$ = 30.  Two calibrators
present in images from both nights passed this constraint, 
2MASSI J1237367+652600 (Calibrator 1) and
2MASSI J1237372+652536 (Calibrator 2).  These sources have 
2MASS J magnitudes of 15.07$\pm$0.05 and 15.34$\pm$0.06, respectively, and
are indicated on Figure 5.

\placefigure{fig-6}

Relative magnitudes
were then computed for 2MASS 1237+6526 using these two 
calibrator stars.  The resulting time series is shown for both 
nights in Figure 6.  The mean relative magnitude has been
subtracted off in both datasets.  Time series for the differenced calibrator
magnitudes ($m_i^{(1)} - m_i^{(2)}$) are also shown (offset) for comparison.
Scatter in the relative magnitudes of
2MASS 1237+6526 was 0.020 and 0.014 mag on May 19 and 20,
respectively, less than the formal
photometric uncertainties.
Corresponding ${\chi}_r^2$ values are 0.505 and 0.314, 
which are greater than those of the
calibrator stars (0.271 and 0.237, respectively),
but nevertheless reject variability (at levels above our
detection threshold) at the $>$ 99\% confidence level.

\placefigure{fig-7}

To examine to possibility of a modulated signal
(e.g., multiple eclipsing events or geometric modulation), 
we constructed periodograms for each night's data using the formalism
of \citet{lom76} and \citet{sca82} as summarized in \citet{pre92}.
Given the duration of the observations, we are sensitive
to periods ranging from roughly 0.1--1.3 hr on each night.  
The computed periodograms (oversampled by a
factor of 10) for both 2MASS 1237+6526 (black lines)
and calibrator stars (grey lines) are shown in
Figure 7.  May 19 and 20 data are identified by solid and dashed lines,
respectively.  
None of the peaks in the 2MASS 1237+6526 periodograms are significant, as they
do not exceed the 50\% false alarm probability level
and are roughly equal in magnitude to insignificant peaks seen 
in the calibrator periodograms.

Variability over two nights also appears to be absent.  
There is a 0.024 mag offset in the
mean relative magnitudes between
the two nights; this is greater than the difference between the
mean calibrator values
(0.003 mag), but comparable to the photometric uncertainties. 
We compute ${\chi}_r^2$ = 0.593 over both nights, again rejecting
variability at the $>$ 99\% confidence level.  Because of the large temporal
gap in the time series data, periodicity analysis requires more 
sophisticated algorithms (e.g., the CLEAN algorithm; Roberts, L{\'{e}}har,
\& Dreher 1987); however, given the lack of statistical evidence for
variability, further analysis appears to be unwarranted. 

Given the null results from these statistical tests, we conclude 
that there is no short-term variability in 2MASS 1237+6526 
at the $\pm$0.025 mag level over periods of 0.1--2.5 hours and no
evidence of periodicity for periods of up to 1.3 hours.  There 
also appears to be no compelling
indication of variability at this level for periods of up to
$\sim$ 14 hours.

\section{The Binary Hypothesis Revisited}

In order to determine what constraints our null variability results 
impose on the interacting binary hypothesis, 
we have updated the basic theory described in \citet{me00b} by
comparison
with theoretical models from \citet{bur97}.  Our observables
are (1) a period limit from the duration of monitoring observations, 
(2) a magnitude limit at which we are sensitive to variability, and
(3) the properties of the primary.  We therefore seek to derive a
relation $\Delta$J = $f(p,i;M_1)$, where $p$ and $i$ are the
period and inclination of the system, and $M_1$ the mass of the primary.
For this discussion, we normalize masses to 
M$_{Jup}$ and distances to R$_{Jup}$, and assume a circular orbit
and negligible angular momentum loss (e.g., to gravitational radiation
or disk interaction).
The period (in hours) is determined from the standard Keplerian relation:
\begin{equation}
p^2 = 4{\pi}^2\frac{a^3}{G(M_1+M_2)} = 8.79\frac{a^3}{M_1(1+q)} 
\label{eq:per}
\end{equation}
in Jupiter-based units.
In the case of an interacting binary,
the secondary fills its Roche lobe, and hence its radius $R_2$
is determined by
\begin{equation}
R_2 = R_L = a\frac{0.49q^{\frac{2}{3}}}{0.6q^{\frac{2}{3}} + \ln{(1+q^{\frac{1}{3}})}}.
\label{eq:roche}
\end{equation}
\citep{egg83}.  Brown dwarf radii, however, are fairly constant from 5--70 
M$_{Jup}$ \citep{bur93}, and are well-described by the relation 
\citep{zap69,ste91}:
\begin{equation}
R = {\alpha}_1\frac{{\pi}M^{-\frac{1}{3}}}{(1+1.8M^{-\frac{1}{2}})^{\frac{4}{3}}},
\label{eq:rm}
\end{equation}
where ${\alpha}_1$ is an age-dependent scale factor
(essentially a correction to partial degeneracy) chosen
to match the interior models
of \citet{bur97}.  Because $R_2$ depends only on the primary mass
and mass ratio $q$, both $a$ and $p$ are functions of M$_1$ and $q$ exclusively.
Therefore, these two parameters define the dynamics of the system.

\placefigure{fig-8}

To determine the observable
change in J-band magnitude caused by a nearly edge-on
eclipse, we assumed a simple 
geometry of overlapping spheres along the line of sight, as diagrammed in 
Figure 8.  In general, because 
dlnR/dlnM $\approx -\frac{1}{3} <$ 0 for brown dwarfs, the lower-mass
secondary has a larger radius ($R_2 > R_1$) for
$M_2 >$ 5 M$_{Jup}$.  The relative area 
of the primary, $A$, obscured by the secondary during maximal eclipse is:
\begin{equation}
{\pi}A =  {\theta}_1 - 0.5\sin{2{\theta}_1} 
 +  (\frac{R_2}{R_1})^2({\theta}_2 - 0.5\sin{{\theta}_2})
\label{eq:area}
\end{equation}
where
\begin{equation}
R_1\sin{{\theta}_1} =  R_2\sin{{\theta}_2}
\label{eq:r1r2}
\end{equation}
\begin{equation}
R_1\cos{{\theta}_1} +  R_2cos{{\theta}_2} = a\cos{i},
\label{eq:r1r2a}
\end{equation}
with ${\theta}_1$ and ${\theta}_2$ defined in Figure 8.  These
equations allow us to uniquely
determine $A$
in terms of $M_1$, $q$, and $i$.  
Note that no eclipse occurs when $\cos{i} > (R_1+R_2)/a$, while
a complete eclipse occurs for $\cos{i} < (R_2-R_1)/a$.

The observed magnitude difference 
at maximal eclipse is then
\begin{equation}
{\Delta}J  \begin{array}[t]{l}
 = -2.5log_{10}(\frac{F_1(1-A) + F_2}{F_1 + F_2}) \\
 = 2.5log_{10}(\frac{A}{1+{\beta}}), \\
 \end{array}
\label{eq:delj1}
\end{equation}
where ${\beta} = F_2/F_1$ is the flux ratio of the secondary to primary
at J-band.  We can determine the change in J-band magnitude
as a function of T$_{eff}$ empirically from the two known
companion T dwarfs, Gliese 229B \citep{nak95} and
Gliese 570D \citep{me00a}.  The former object is a T6.5 V with M$_J$ = 
15.52$\pm$0.06 \citep{leg99} and T$_{eff}$ $\sim$ 950 K 
\citep{all96,mar96,tsu96b}; the latter object is a T8 V with M$_J$ = 
16.47$\pm$0.07 \citep{me00a} and T$_{eff}$ $\sim$ 800 K \citep{geb01a}.
These values yield the very approximate relation dJ/dT = $-$0.0063 mag/K,
which may only be accurate over a limited T$_{eff}$ range\footnote{This
rough relation also ignores the effects of
gravity and metallicity, which could be substantial.}.
Interior models from \citet{bur97} show a power law relation between 
T$_{eff}$ and mass which is roughly dependent on age (${\tau}$).  
A fit to the models
yields the relations:
\begin{equation}
\frac{dT}{dM} = \left\{ \begin{array}{ll}
 88.7 M^{-0.43} & {\tau} = 0.5\ Gyr \\
 73.4 M^{-0.44} & {\tau} = 1.0\ Gyr \\
 45.4 M^{-0.44} & {\tau} = 5.0\ Gyr \\
 \end{array}
 \right. K/M_{Jup}.
\label{eq:dtdm}
\end{equation}
Combining Equation~\ref{eq:dtdm} with the expression for dJ/dT 
yields the secondary flux ratio:
 \begin{equation}
-2.5log_{10}{\beta}  \begin{array}[t]{l}
     = \frac{dJ}{dT}\frac{dT}{dM}{\Delta}M \\
     = \frac{dJ}{dT}\frac{dT}{dM}M_1(q-1) \\
     = {\alpha}_2 {\times} (1-q) \\
 \end{array}
\label{eq:sigma}
\end{equation}
where ${\alpha}_2$ is a constant dependent on age and 
primary mass.  
We have assumed that 2MASS 1237+6526, a T6.5 V similar to Gliese 229B, has a
T$_{eff}$ $\sim$ 950 K, and masses 23, 30, and 55 M$_{Jup}$ for ${\tau}$ = 0.5,
1.0, and 5.0 Gyr, respectively \citep{bur97}.  
The various parameters
are summarized in Table 3.

\placetable{tbl-3}
\placefigure{fig-9}

Using the above equations,
we have calculated
the expected magnitude drop ($\Delta$J$_{ecl}$)
at minimum light for an eclipsing, interacting brown dwarf
binary with T$_{primary}$ = 950 K, 0.5 $\leq {\tau} \leq$ 5.0 Gyr,
0 $\leq q \leq$ 1, and
40$\degr \leq i \leq$ 90$\degr$.  Results are shown in Figure 9
for $\tau$ = 0.5 (solid line), 1.0 (dot-dashed line), and 5.0 (dashed line)
Gyr, and (from top to bottom) $i$ = 90$\degr$, 80$\degr$, 70$\degr$,
and 60$\degr$.  Only values of $\Delta$J$_{ecl}$ for which $q <$ 0.63
and M$_2$ $>$ 5 M$_{Jup}$
are displayed, corresponding to the limits for 
sustained mass loss \citep{me00b}.
Significant drops in the observed flux are predicted by
this model,
particularly for $i$ $\geq$ 80$\degr$, where fluxes at maximum eclipse
over 0.7 mag fainter than the nominal brightness are possible
over the full range of periods shown.  
These flux deviations
are due primarily to the substantial brightness ratio
between the secondary and primary, which is particularly 
important at 
high inclination angles where obscuration is most complete.  
At lower inclination angles, the limiting factor is the
orbital separation, which is larger
for lower mass secondaries.  No eclipse variations are predicted to occur
for $i \lesssim 50\degr$ over the range of ages examined.
Note that
significant systematic errors may be present for the lowest
mass ratios (i.e., low secondary T$_{eff}$)
due to the rough estimate of $\beta$ in 
Equation~\ref{eq:sigma}; however, our observations
generally sample only the highest mass ratios due to the period
cutoff.

The predicted $\Delta$J$_{ecl}$ from these models are
generally much larger than our observational limits (indicated by
the hatched region in Figure 8).  The constraints
imposed by the non-detection of variability 
are summarized in Table 3 for $i$ $>$ 60$\degr$;
in this case, $q$ $>$ 0.70, 
0.53, and 0.37, corresponding to secondary masses M$_2$ $>$ 16, 16, 
and 20 M$_{Jup}$ for ${\tau}$ = 0.5, 1.0, and 5.0 Gyr, respectively.
Note that for the 0.5 Gyr case, our  
mass ratio limit exceeds the minimum
$q$ required to sustain mass loss.  The lowest inclination angle
eclipses that could have been detected by our observations
for ${\tau}$ = 1.0 and 5.0 Gyr are
$i_{min}$ = 54$\degr$ and 56$\degr$, with 
corresponding minimum secondary
masses of 20 and 34 M$_{Jup}$, respectively.

\section{Discussion}

Based on the results of the monitoring observations, 
we can confidently rule out 
an interacting 
binary system for 2MASS 1237+6526 for $i$ $>$ 60$\degr$ and
M$_2$ $>$ 20 M$_{Jup}$.  We cannot rule out a less inclined
system, although such a geometry would fail
to explain the apparent line shift seen in our 2000 March spectral data.
A compromise solution would be a moderately inclined 
system with a massive primary and large mass ratio.  Conclusive
synthesis of the spectroscopic and photometric data, however, requires
verification of H$\alpha$ line shift.

Other emission mechanisms should also be considered.
2MASS 1237+6526 could be a young, and hence very low mass, brown dwarf
that is still accreting material from a circum(sub)stellar disk.
If this object were in a weak-lined T-Tauri (WLTT) phase, 
its H$\alpha$ line could arise from a small, optically thin
accretion boundary layer,
consistent with our spectral and photometric observations.
Assuming a typical WLTT age (30--100 Myr; 
Hartmann, Kenyon, \& Hartigan 1993) and T$_{eff}$ $\sim$ 950 K, 
the evolutionary models of
\citet{bur97} predict a mass of 3--12 M$_{Jup}$, below the Deuterium-burning
limit.  Note that
2MASS 1237+6526 is not associated with any known star-forming region, 
although it could have been ejected
from its nascent cluster \citep{rep01}.  The absence of the
1.25 $\micron$ K I doublet in this object (I.\ McLean, priv.\ comm.) marginally
supports the 
possibility of it being a low-gravity source, although we cannot rule out
a purely temperature effect \citep{me01a}.

Our follow-up observations of 2MASS 1237+6526 have enabled us to 
rule out flaring as a viable means of emission, and have placed
significant constraints on the geometry and membership of an interacting
binary system.
Parallax (to determine absolute brightness) and space motion 
measurements are clearly required to further characterize
this object.
Furthermore, the apparent velocity shift described in $\S$2.4, which
may prove to be a vital clue to the origin of emission in 2MASS 1237+6526,
requires
confirmation and more detailed, higher-resolution investigation.  
The nature of this enigmatic brown dwarf may hopefully be revealed
with further observations.

\acknowledgements

A.\ J.\ B. would like to thank Telescope Operators
Terry Stickel (Keck) and Barrett ``Skip'' Staples (Palomar)
for their support during the observations presented here.
A.\ J.\ B. also acknowledges useful discussions with 
C.\ A.\ L.\ Bailer-Jones, D.\ Koerner, M.\ Marley, I.\ McLean, and
M.\ R.\ Zapatero Osorio during the 
preparation of the manuscript, and valuable comments from our anonymous
referee.
A.\ J.\ B., J.\ D.\ K., and J.\ E.\ G.\ acknowledge the support 
of the Jet Propulsion
Laboratory, California Institute of Technology, which is operated under
contract with the National Aeronautics and Space Administration.  
A.\ J.\ B.\ also acknowledges support provided by NASA through a 
Hubble fellowship grant from the Space Telescope Science Institute,
which is operated by the Association of Universities for Research in Astronomy,
Incorporated, under NASA contract NAS5-26555.
J.\ L.\ acknowledges support from NASA JPL grant 961040NSF.
Portions of the data presented herein were obtained at the 
W.\ M.\ Keck Observatory which is operated as a scientific partnership 
among the California Institute of Technology, the University of California,
and the National Aeronautics and Space Administration. The Observatory was 
made possible by the generous financial support of the W.\ M.\ Keck Foundation.
This publication makes use of data from the Two Micron
All Sky Survey, which is a joint project of the University of
Massachusetts and the Infrared Processing and Analysis Center,
funded by the National Aeronautics and Space Administration and
the National Science Foundation.

\clearpage

\figcaption[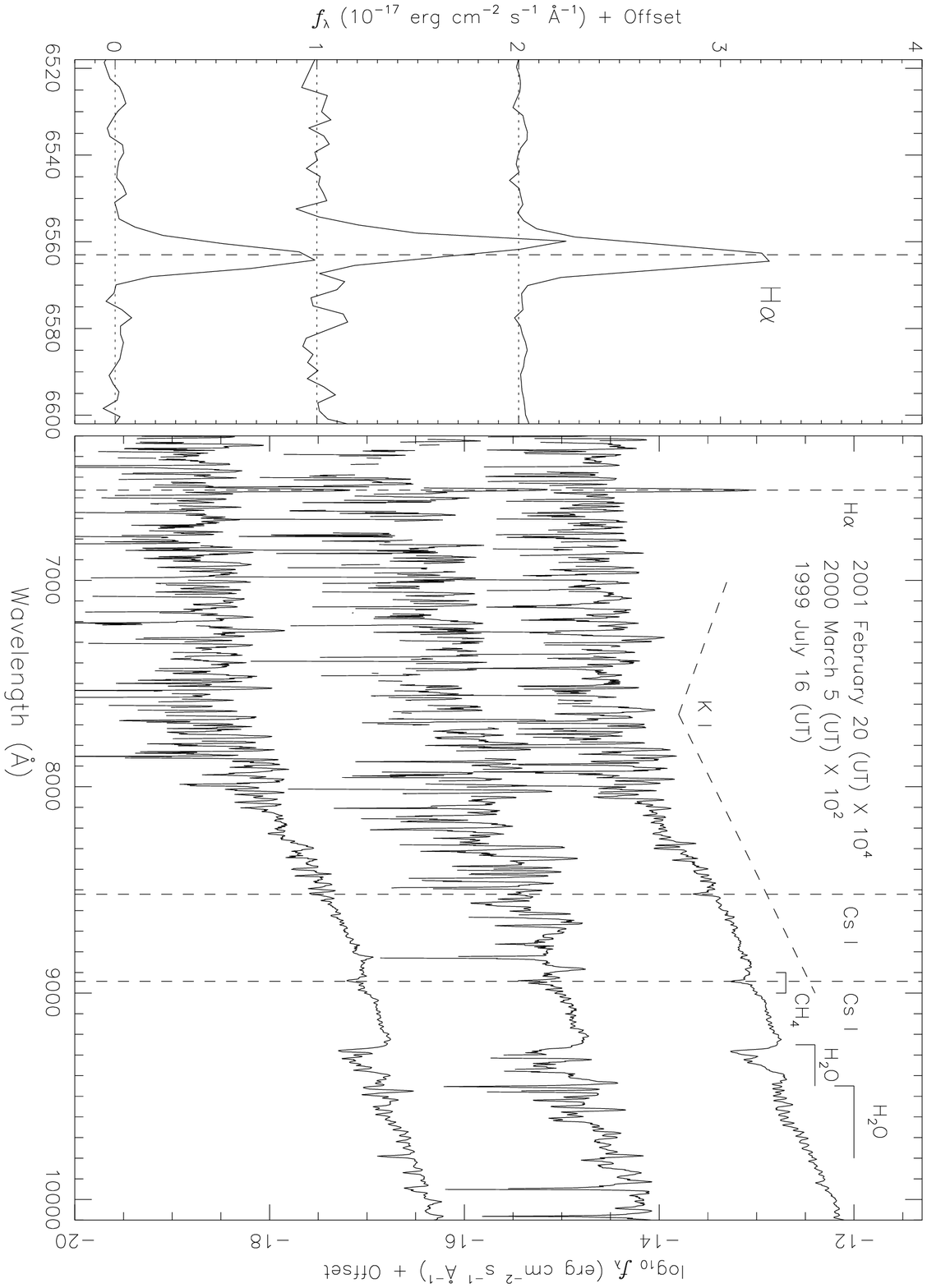]{Optical spectra of 2MASS 1237+6526 
obtained using LRIS.  Observations from 1999 July 16,
2000 March 5, and 2001 February
20 (UT) are shown, and are scaled by the correction factors described in
$\S$2.2.  The right panel
shows full red optical spectra from 6300--10100 {\AA}, with flux plotted
on a logarithmic scale to enhance features; spectra are offset by a
constant factor.  Observed bands of H$_2$O and CH$_4$, and lines
of Cs I and H$\alpha$ are labelled, along with the pressure-broadened
K I doublet.  The left panel shows a closeup of the H$\alpha$ line on a
linear scale, with data offset by an additive constant (dotted lines).
\label{fig-1}}

\figcaption[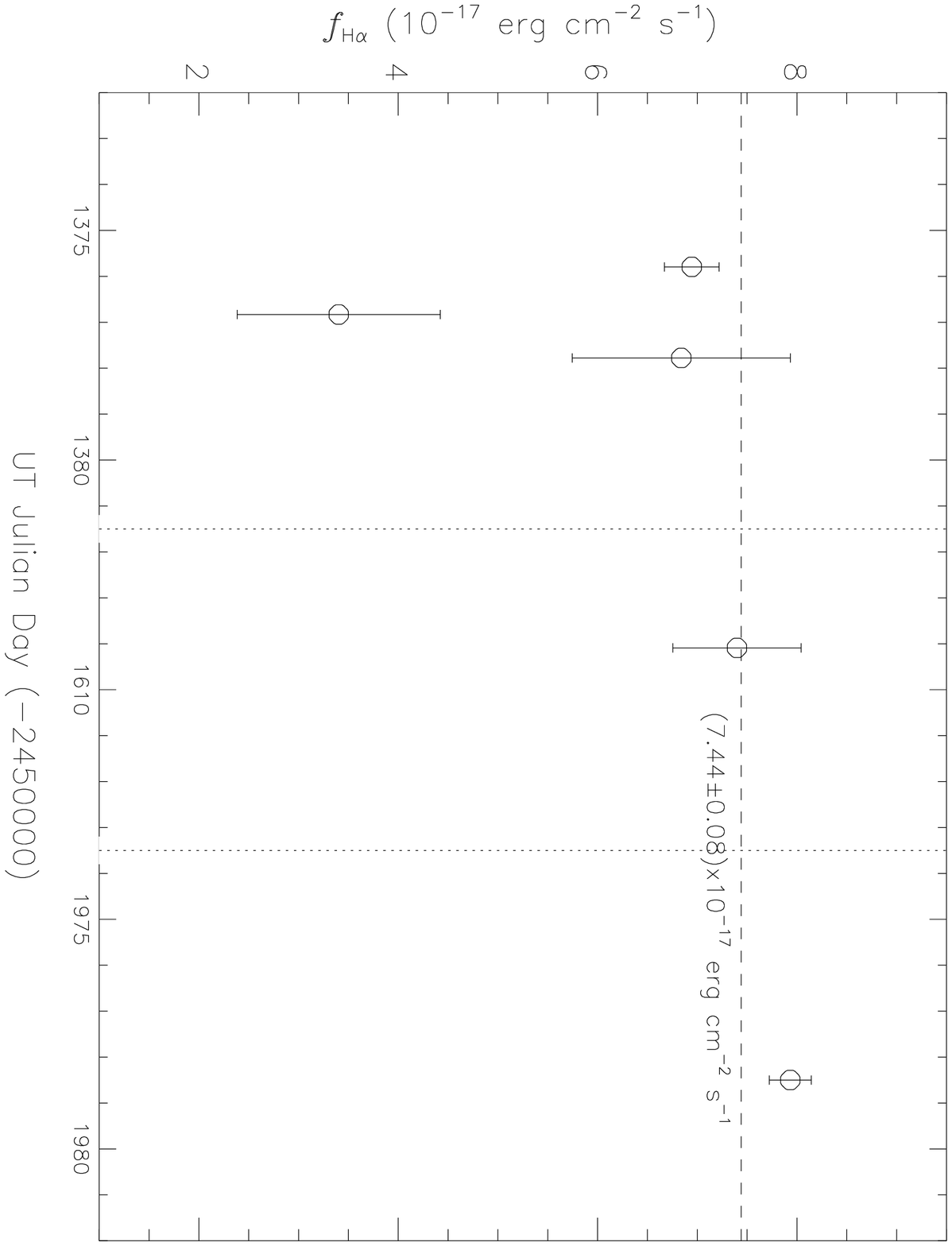]{Emitted H$\alpha$ flux from
2MASS 1237+6526 versus UT time.
Circles plot corrected line flux values, 
scaled by matching
flux data in the 8000-9500 {\AA} region to observations made on
1999 July 16 (UT).  The mean emission strength 
$f_{H{\alpha}}$ = (7.44$\pm$0.08)$\times$10$^{-17}$ erg cm$^{-2}$ s$^{-1}$
is indicated by the dashed line.
\label{fig-2}}

\figcaption[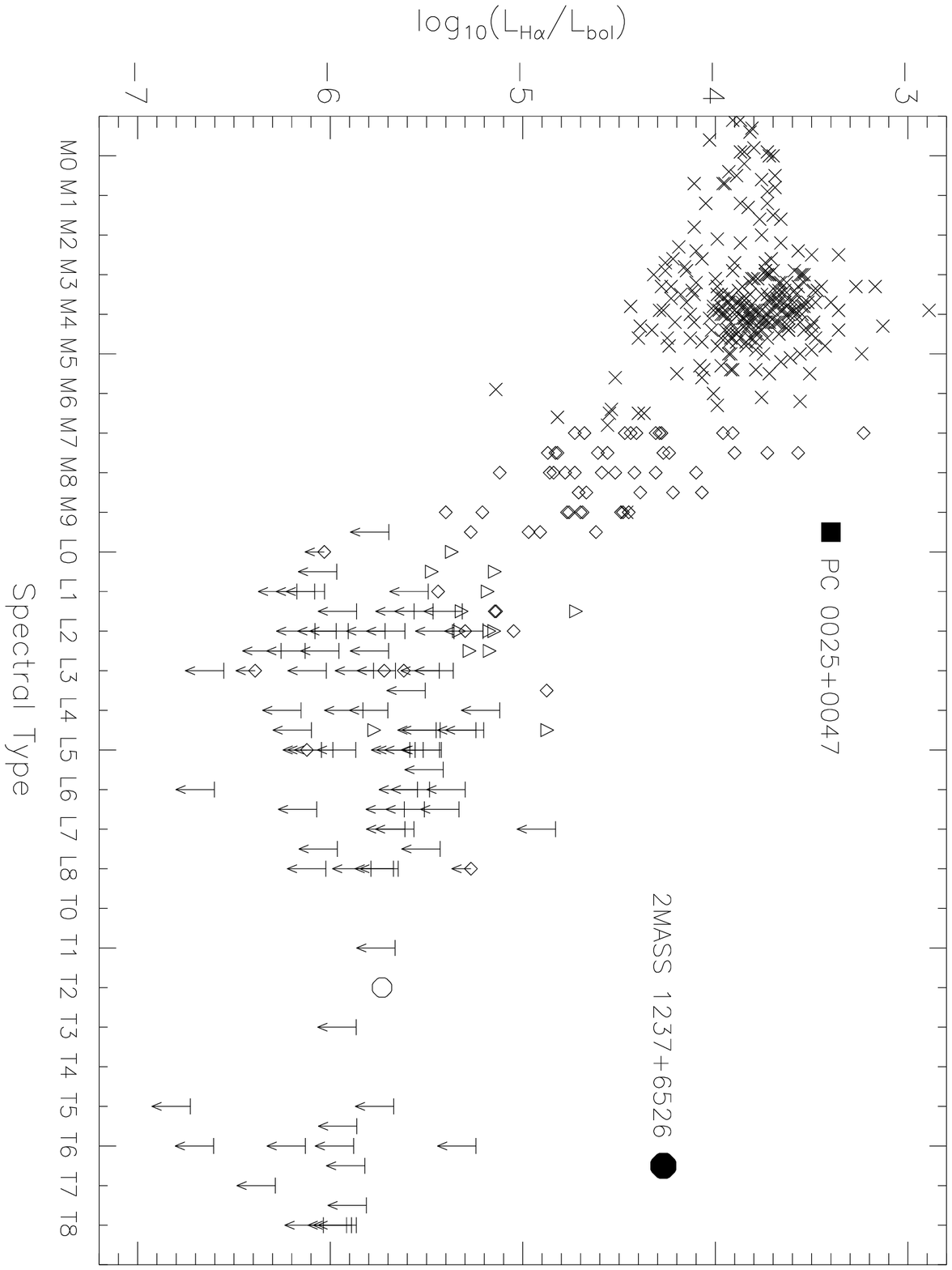]{Relative H$\alpha$ luminosity
versus spectral type for cool
dwarfs.  The mean relative luminosity for 2MASS 1237+6526 is indicated
by the solid circle, while values calculated from late-M, L, and T dwarf
spectra are shown as open symbols (detected) or arrows 
(nondetection upper limits).
Also shown are data for M and L dwarfs
from \citet[crosses]{haw96} and \citet[diamonds]{giz00}, and 
for PC 0025+0047 \citep[solid square]{sch91}.  T spectral types are
from \citet{me01a}.
\label{fig-3}}

\figcaption[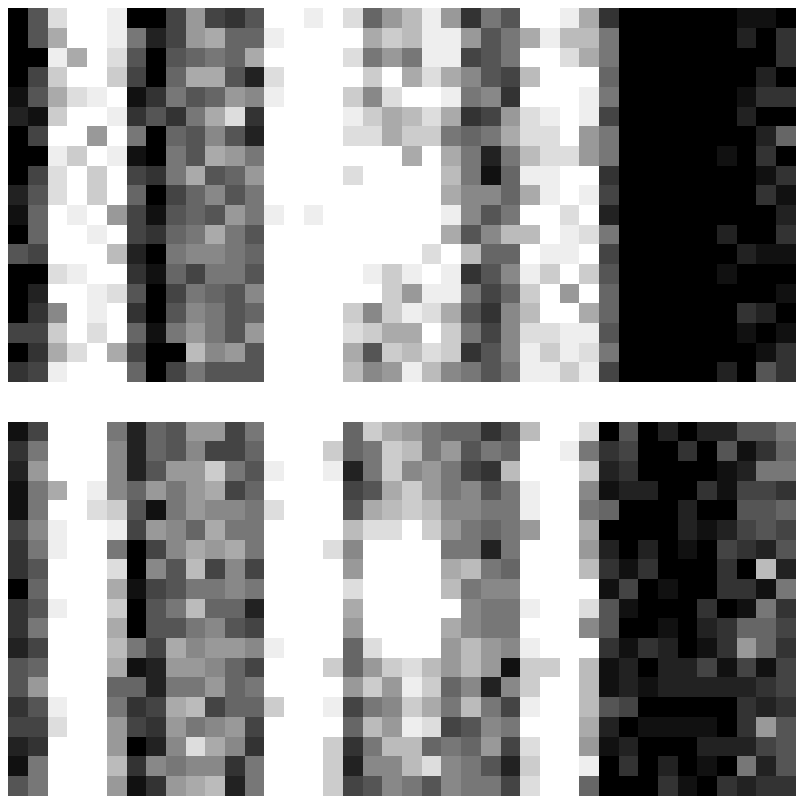]{Raw LRIS
data of 2MASS 1237+6526 from 2000 March 5 (top)
and 2001 February 20 (UT, bottom) in the 6563 {\AA} region.  Vertical
bands are telluric OH emission lines, while the bright spots arise
from H$\alpha$ emission in 2MASS 1237+6526.
\label{fig-4}}

\figcaption[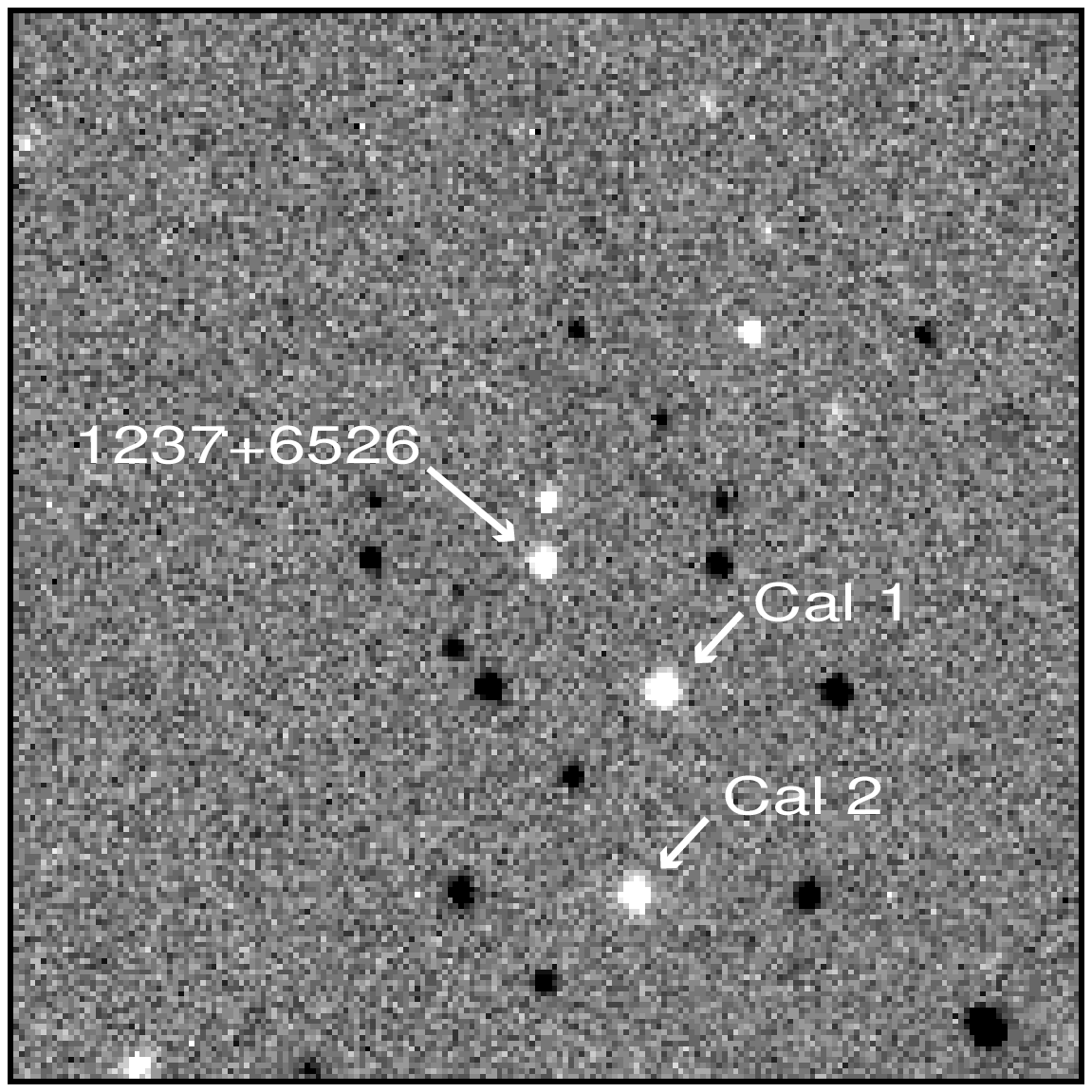]{Combination of five, 45 sec J-band images
of the 2MASS 1237+6526 field obtained on 2000 May 19 (UT).  2MASS 1237+6526
is indicated, as are the two comparison stars Cal 1 and Cal 2.
The field shown is
$2\arcmin{\times}2\arcmin$, oriented with North up and East toward the left.
\label{fig-5}}

\figcaption[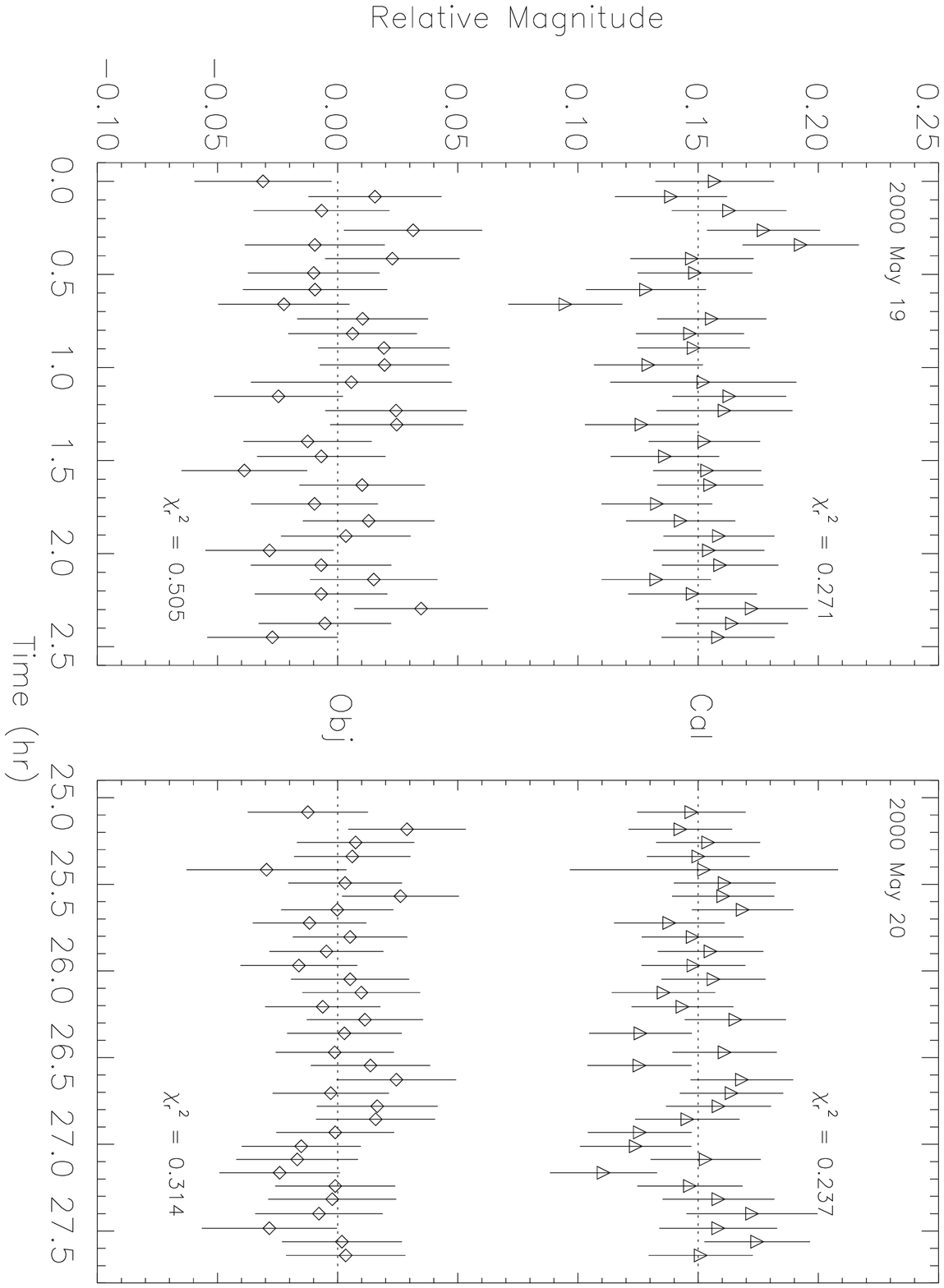]{Relative magnitude time series for 
2MASS 1237+6526 (diamonds) and comparison stars (triangles).  Data for
2000 May 19 and 20 (UT) are shown separately.  The
mean relative magnitudes for each night have been subtracted, with 
datasets offset for clarity. 
\label{fig-6}}

\figcaption[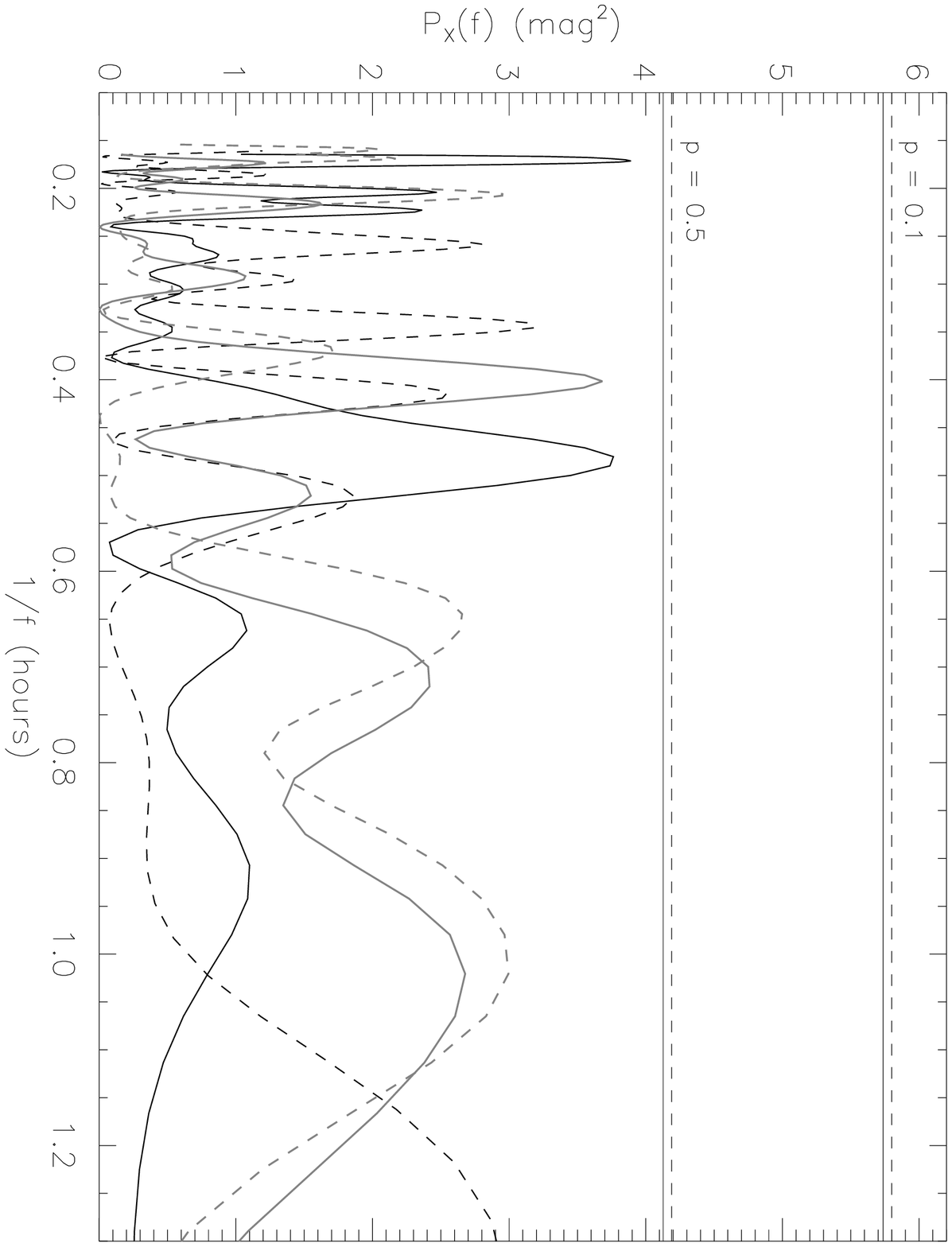]{Periodograms of 
monitoring time series data.  Black and grey lines trace the periodograms of
2MASS 1237+6526 and comparison stars, respectively, while solid and dashed 
lines identify data from 2000 May 19 and 20 (UT).  False alarm probability
limits of 50\% and 10\% are indicated. 
\label{fig-7}}

\figcaption[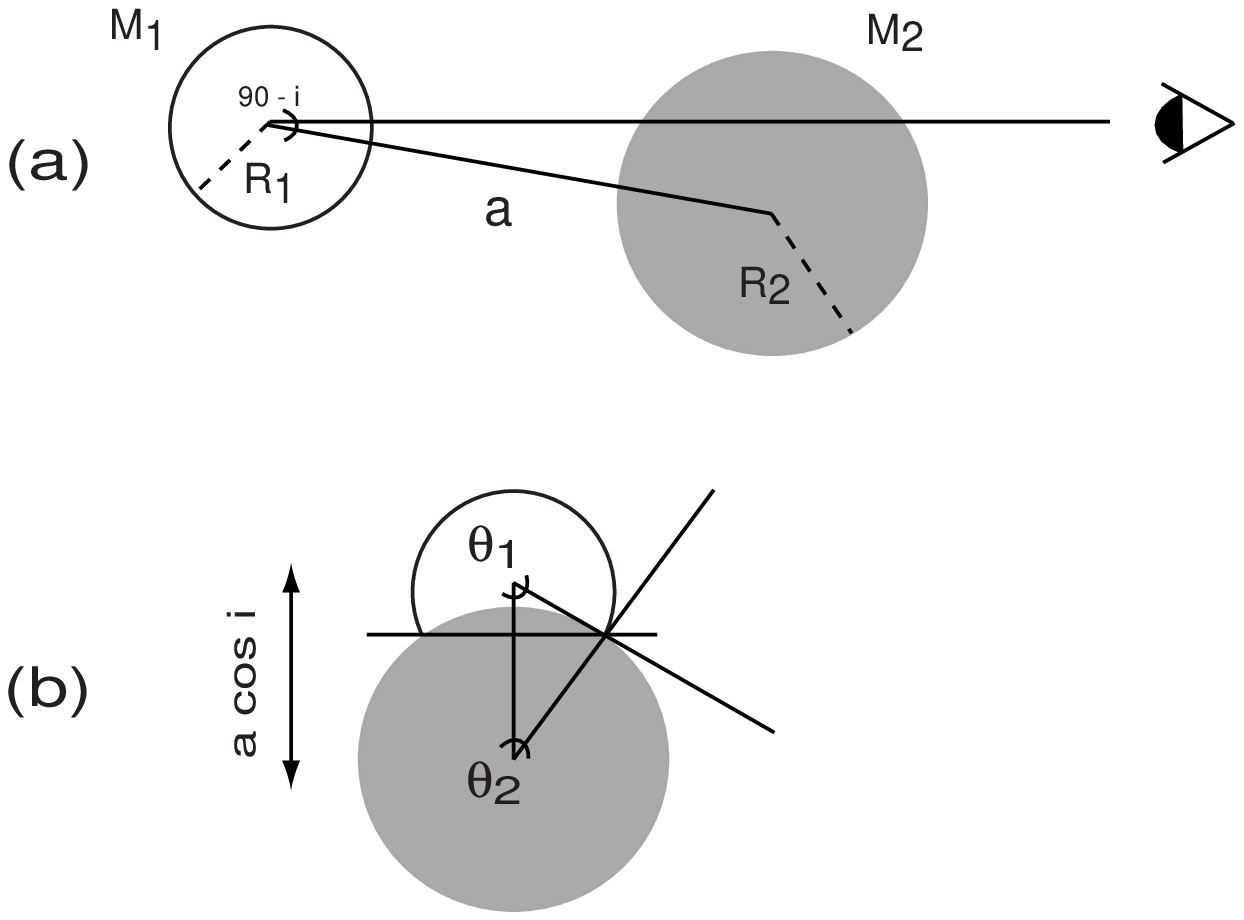]{Geometry of the interacting binary model.
(a) Orientation along the line-of-sight; $a$ is the binary
separation, $i$ the inclination, $M_1$ and $M_2$ the masses
of the primary and secondary, and $R_1$ and $R_2$ their respective
radii. (b) Face-on orientation, showing the relevant angles,
${\theta}_1$ and ${\theta}_2$, used
to compute the obscured area of the primary.
\label{fig-8}}

\figcaption[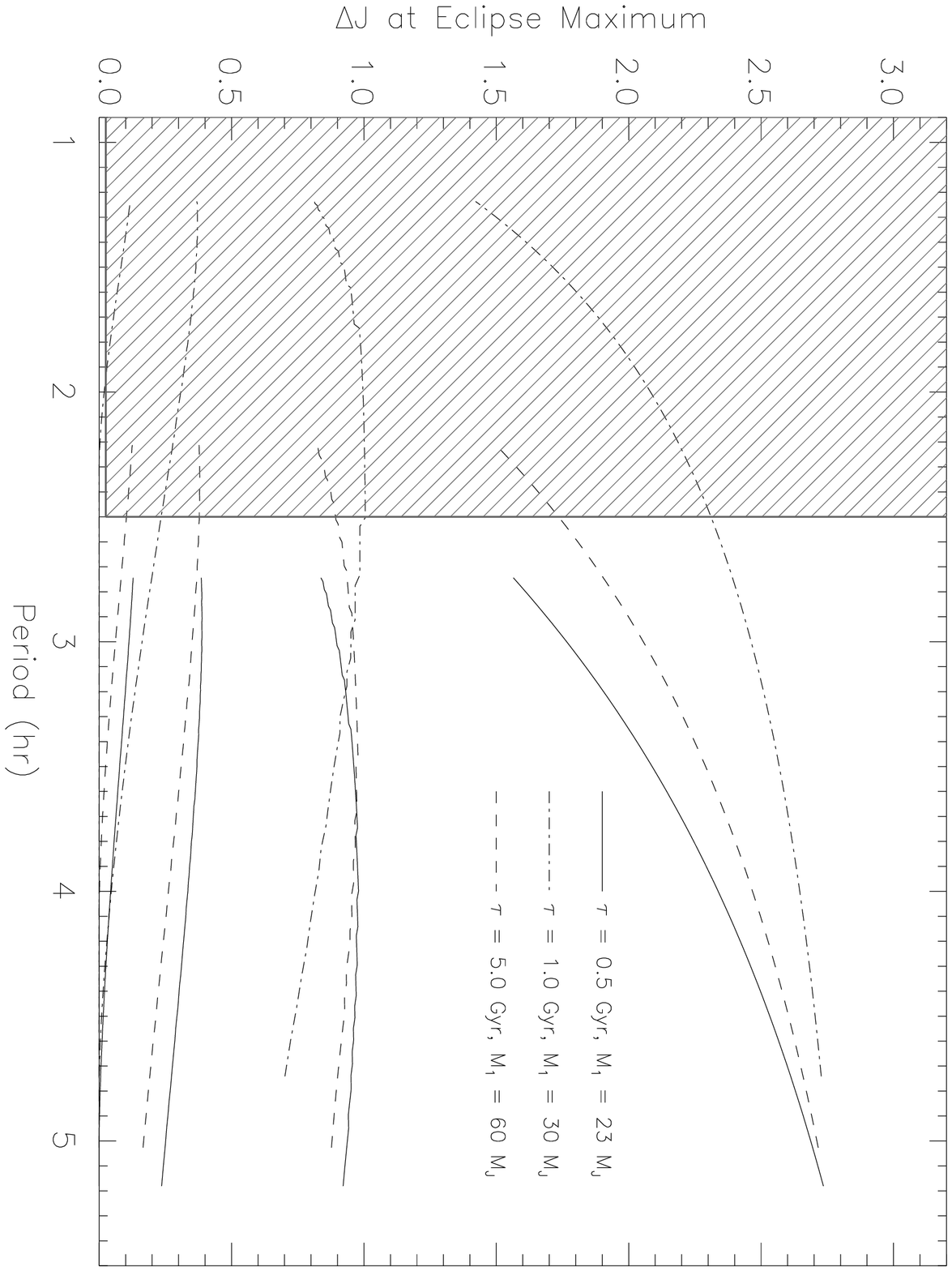]{Constraints on the binary hypothesis.
Predicted $\Delta$J$_{ecl}$ during maximal eclipse is plotted versus period for
$q <$ 0.63; M$_2$ $>$ 5 M$_{Jup}$; 
$\tau$ = 0.5 (solid line), 1.0 (dot-dashed line), and 5.0 (dashed line)
Gyr; and (from top to bottom) $i$ = 90$\degr$, 80$\degr$, 70$\degr$,
and 60$\degr$.  The hatched area covers the parameter space probed by
the J-band monitoring observations.
\label{fig-9}}

\clearpage

\begin{deluxetable}{lcccc}
\tabletypesize{\scriptsize}
\tablenum{1}
\tablewidth{0pt}
\tablecaption{H$\alpha$ Line Strengths. \label{tbl-1}}
\tablehead{
\colhead{} &
\colhead{} &
\colhead{Uncorrected} &
\colhead{Corrected} &
\colhead{Corrected} \\
\colhead{UT Date} &
\colhead{t$_{int}$ (s)} &
\colhead{$f_{H\alpha}$\tablenotemark{a}} &
\colhead{$f_{H\alpha}$\tablenotemark{a}} &
\colhead{$\log$ (L$_{H\alpha}$/L$_{bol}$)} \\
\colhead{(1)} &
\colhead{(2)} &
\colhead{(3)} &
\colhead{(4)} &
\colhead{(5)} 
}
\startdata
1999 July 16 & 4800 & {\phn}6.9$\pm$0.3 & 6.9$\pm$0.3 & $-$4.3 \\
1999 July 17 & 1800 & {\phn}1.6$\pm$0.5 & 3.4$\pm$1.0 & $-$4.6 \\
1999 July 18\tablenotemark{b} & 1800 & {\phn}8.4$\pm$1.4 & 6.8$\pm$1.1 & $-$4.3 \\
2000 March 5 & 1800 & {\phn}2.9$\pm$0.3 & 7.4$\pm$0.6 & $-$4.3 \\
2001 February 20 & 2400 & 10.6$\pm$0.3 & 7.9$\pm$0.2 & $-$4.2 \\
\enddata
\tablenotetext{a}{In units of 10$^{-17}$ erg cm$^{-2}$ s$^{-1}$.}
\tablenotetext{b}{Data obtained in the 3800--8600 {\AA} regime.}
\end{deluxetable}

\begin{deluxetable}{ccccl}
\tabletypesize{\scriptsize}
\tablenum{2}
\tablewidth{0pt}
\tablecaption{Log of Imaging Observations. \label{tbl-2}}
\tablehead{
\colhead{UT Date} &
\colhead{UT Time} &
\colhead{Total t (hr)} &
\colhead{No. Images} &
\colhead{Conditions} \\
\colhead{(1)} &
\colhead{(2)} &
\colhead{(3)} &
\colhead{(4)} &
\colhead{(5)} 
}
\startdata
19 May 2000 & 05 31 26 -- 08 01 55 & 2.51 & 155 & hazy and clear \\
20 May 2000 & 06 36 28 -- 09 13 25 & 2.62 & 165 & hazy with some scattered cirrus 
\enddata
\end{deluxetable}

\begin{deluxetable}{cccccccc}
\tabletypesize{\scriptsize}
\tablenum{3}
\tablewidth{0pt}
\tablecaption{Parameters for the 2MASS 1237+6526 Binary Model. \label{tbl-3}}
\tablehead{
\colhead{$\tau$ (Gyr)} &
\colhead{${\alpha}_1$\tablenotemark{a}} &
\colhead{M$_1$ (M$_{Jup}$)} &
\colhead{${\alpha}_2$ (mag)\tablenotemark{a}} &
\colhead{$q_{max}$\tablenotemark{b}} &
\colhead{$M_{2,max}$ (M$_{Jup}$)\tablenotemark{b}} \\
\colhead{(1)} &
\colhead{(2)} &
\colhead{(3)} &
\colhead{(4)} &
\colhead{(5)} &
\colhead{(6)} 
}
\startdata
0.5 & 1.39 & 23 & 3.5 & 0.70\tablenotemark{c} & 16  \\
1.0 & 1.36 & 30 & 3.3 & 0.53 & 16  \\
5.0 & 1.30 & 55 & 3.1 & 0.37 & 20  \\
\enddata
\tablenotetext{a}{Scaling factor determined by comparison to the structure
models of \citet{bur97}.}
\tablenotetext{b}{Maximum values permitted by non-detection of variability
for $i$ $>$ 60 $\degr$.}
\tablenotetext{c}{This value is greater than the maximum $q$ = 0.63 required
to sustain mass loss.}
\end{deluxetable}

\clearpage
\onecolumn
\setcounter{figure}{0}

\begin{figure}
\epsscale{0.9}
\plotone{Burgasser.fig1.ps}
\caption{}
\end{figure}
\clearpage

\begin{figure}
\epsscale{0.9}
\plotone{Burgasser.fig2.ps}
\caption{}
\end{figure}
\clearpage

\begin{figure}
\epsscale{0.9}
\plotone{Burgasser.fig3.ps}
\caption{}
\end{figure}
\clearpage

\begin{figure}
\epsscale{0.9}
\plotone{Burgasser.fig4.eps}
\caption{}
\end{figure}
\clearpage

\begin{figure}
\epsscale{0.9}
\plotone{Burgasser.fig5.eps}
\caption{}
\end{figure}
\clearpage

\begin{figure}
\epsscale{0.9}
\plotone{Burgasser.fig6.ps}
\caption{}
\end{figure}
\clearpage

\begin{figure}
\epsscale{0.9}
\plotone{Burgasser.fig7.ps}
\caption{}
\end{figure}
\clearpage

\begin{figure}
\epsscale{0.9}
\plotone{Burgasser.fig8.eps}
\caption{}
\end{figure}
\clearpage

\begin{figure}
\epsscale{0.9}
\plotone{Burgasser.fig9.ps}
\caption{}
\end{figure}
\clearpage

\end{document}